\documentstyle[12pt,epsf,epsfig,aps]{revtex}
\tighten
\begin{document}
\title{Homogeneous spacelike singularities inside spherical black holes}
\author{Lior M. Burko}
\address{Department of Physics, Technion---Israel Institute of Technology, 
32000 Haifa, Israel}
\date{\today}
\maketitle

\begin{abstract}

Recent numerical simulations have found that the Cauchy horizon inside
spherical charged black holes, when perturbed nonlinearly by a self-gravitating, 
minimally-coupled,  massless, spherically-symmetric scalar field, turns into a null weak
singularity which focuses monotonically to $r=0$ at late times, where the
singularity becomes spacelike. 

Our main objective is to study this spacelike singularity. 
We study analytically the spherically-symmetric Einstein-Maxwell-scalar
equations asymptotically near the singularity. 
We obtain a series-expansion solution for the metric functions
and for the scalar field near $r=0$ under the simplifying assumption of
homogeneity. Namely, we neglect spatial derivatives and keep only temporal derivatives.  
We find that there indeed exists a {\em generic} spacelike singularity solution
for these equations (in the sense that the solution depends on enough
free parameters), with similar properties to those found in the numerical
simulations. This singularity is strong in the Tipler sense, namely,
every extended object would inevitably be crushed to zero volume. In this
sense this is a similar singularity to the spacelike singularity inside
uncharged spherical black holes. On the other hand, there are some
important differences between the two cases. Our model can also be
extended to the more general inhomogeneous case.

The question of whether the same kind of singularity evolves in more
realistic models (of a spinning black hole coupled to gravitational
perturbations) is still an open question.

\end{abstract}

\section{Introduction}\label{sec1}

The singularity theorems of Hawking and Penrose \cite{bur-haw73} predict the 
occurrence of spacetime singularities inside black holes under very 
plausible assumptions. However, they tell us nothing about the geometrical 
and physical nature and properties of these singularities. 
The Kerr black hole, which describes the exterior of realistic black holes, 
fails to describe adequately their interiors. In the Kerr solution the 
singularity has the shape of a ring, and it is timelike. Consequently, 
it violates the strong cosmic censorship hypothesis of 
Penrose \cite{bur-pen69}. However, 
the Kerr solution is unstable, in the sense that the Cauchy horizon 
is converted into a spacetime singularity. This singularity is null and 
weak \cite{bur-ori92}. 
Namely, infalling extended observers would experience only finite 
tidal distortions at the singularity. Because of the complexities involved 
with the analysis of spinning black holes, a frequently used toy model is 
the spherical charged black hole. Although realistic black holes are not 
expected to be significantly charged, the spherical charged black holes 
share many common properties and similar causal structure with the more 
realistic spinning black holes. With spherical symmetry, non-trivial dynamics 
can be modeled by a scalar field, which has a spherically symmetric 
radiative mode. 
In recent numerical analyses with this toy model (namely, a spherical 
charged black hole perturbed non-linearly by a self-gravitating scalar 
field) it has been shown that the generators of the Cauchy horizon are 
focused at the late parts of the Cauchy horizon \cite{bur-bra95,bur-bur97a}. 
Namely, the value of the area coordinate $r$ decreases slowly and monotonically 
along the Cauchy horizon, until the Cauchy horizon is completely focused, 
i.e., the area coordinate shrinks to zero value at a finite value of the 
affine parameter along the Cauchy horizon. Then, the singularity becomes 
spacelike. (The spacelike singularity was first found numerically by 
Gnedin and Gnedin \cite{bur-gne93}.) 
It is still an open question whether in 
a realistic spinning black hole perturbed non-linearly by 
a realistic physical field (such as gravitational waves) the generators  
of the Cauchy horizon would also focus completely, and create a 
strong spacelike singularity. 
In this paper we shall focus on the spacelike singularities inside spherical black holes 
(both uncharged and charged), and study their properties under the simplifying assumptions of 
spherical symmetry and a minimally-coupled massless scalar field. We shall also assume 
homogeneity and power-law behavior of the metric functions, 
which are justified by recent numerical simulations. 

The spacelike singularity inside a spherically symmetric uncharged black hole 
perturbed by a scalar field was studied by Doroshkevich and Novikov \cite{bur-dor78}, who 
treated the scalar field as a linear perturbation and 
obtained the general solution for the linear scalar field, 
but did not study the backreaction of the scalar field on the geometry. 
Doroshkevich and Novikov 
found that the scalar field diverges logarithmically with $r$ at the 
singularity. Therefore, the energy density of the scalar field diverges 
like an inverse power of $r$, and therefore, acting as a source term 
for the Einstein-Klein-Gordon equations, the scalar field might be expected to 
change the metric functions. The very divergence of the linear 
perturbation analysis indicates that nonlinear effects are expected 
to be important. More recently, 
Krori, Goswami and Das Purkaystha \cite{bur-kro95} considered the same 
problem as Doroshkevich and Novikov \cite{bur-dor78}, but failed to obtain a general 
solution. The regular solution obtained by these authors depends on just 
one arbitrary parameter, while a general solution clearly depends on 
two independent arbitrary parameters (see below). Consequently, the correct 
conclusions to be drawn from a linear analysis are the conclusions of 
Doroshkevich and Novikov \cite{bur-dor78} and not the conclusions of 
Krori, Goswami and Das Purkaystha \cite{bur-kro95}.  
However, with the presence of an electric charge a linear perturbation 
analysis cannot investigate a generic spacelike singularity, as this 
singularity is created by nonlinear effects, i.e., by the nonlinear focusing 
of the generators of the Cauchy horizon. Under linear perturbations there would 
not exist inside a spherical charged black hole a spacelike singularity at all. 
Therefore, to study the spacelike 
singularity inside a spherical charged black hole with a scalar field one 
has to consider the fully nonlinear case. Nonlinear analysis of spacetime 
singularities with a scalar field in the cosmological 
case was done by Belinskii and Khalatnikov \cite{bur-bel73}, 
who found that the scalar field destroys the BKL oscillations and that 
in that case the singularity is monotonic. 
(Interestingly, Belinskii and Khalatnikov showed that if, 
in addition to the scalar field, there were also a vector field, 
the singularity would again be oscillatory.  
The analysis by Belinskii and Khalatnikov \cite{bur-bel73} is markedly different from the 
analysis we shall present below, as we are interested here in spherical symmetry. It turns out 
that the Kasner-metric based Belinskii--Khalatnikov analysis cannot be reconciled with spherical 
symmetry.)

The organization of the paper is as follows. In Section \ref{sec2} we describe the physical 
model we employ. In Section \ref{sec3} we discuss the properties of power-law 
spacelike singularities. In sections \ref{sec4} and \ref{sec5} we study the cases 
of uncharged and charged spherical black holes, correspondingly. 
We show that indeed there exists a generic solution
of the spherically symmetric coupled Einstein-Maxwell-Klein-Gordon equations with a
spacelike singularity, with properties which one actually finds in
numerical simulations. This result strengthens our confidence in the picture of
black hole interiors as described above. In Section \ref{sec6} we 
analyze the strength of the singularity we find. We find that it is strong in the sense of Tipler (whereas
the null Cauchy horizon singularity is weak). We summarize and give some concluding 
remarks in Section \ref{sec7}. 

\section{The physical model}\label{sec2}
\begin{figure}
\input{epsf}
\centerline{\epsfysize 7.0cm
\epsfbox{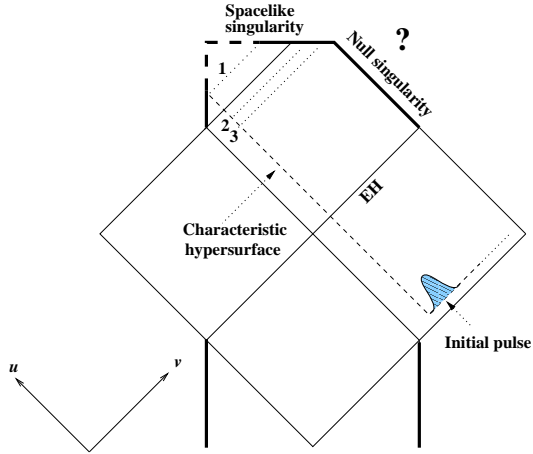}}
\caption{Spacetime diagram for a spherical charged black hole 
with a self-gravitating scalar field. Singularities are denoted by thick lines. 
Prior to the initial hypersurface (dashed) the geometry is 
Reissner-Nordstr\"{o}m. The spacetime is then perturbed by the scalar field. 
The Cauchy horizon is converted into a null singularity. It is still an open 
question whether there is a continuation of the spacetime manifold beyond the 
null singularity, and if so, what the topology and the geometry are. 
The Cauchy horizon is focused to $r=0$, where the singularity is spacelike. 
Our numerical setup does not allow us to investigate the late parts of the 
spacelike singularity, which are denoted by a thick 
dashed line, as they are located beyond the domain of influence of the 
characteristic hypersurface. We investigate the fields along outgoing null 
curves, denoted by $1$, $2$, and $3$.}
\label{fig1}
\end{figure}

We performed numerical simulations of the collapse of a spherically symmetric 
self-gravitating minimally coupled massless scalar field over a pre-existing 
Reissner-Nordstr\"{o}m black hole. Our code is based on free evolution 
and on double-null coordinates and is described at length in 
Ref. \cite{bur-bur97}. The code is stable and converges with second order.  
The numerical set-up is described in Figure \ref{fig1}. Prior to the initial 
hypersurface the geometry is Reissner-Nordstr\"{o}m. Then, at some 
advanced time $v_{0}$ the spacetime is perturbed by a high-amplitude 
self-gravitating spherical massless scalar field of squared-sine shape 
on the outgoing section of the initial hypersurface. The null coordinates 
we use in the numerical simulations are linear with the area coordinate 
$r$ on the initial hypersurface.  

\begin{figure}
\input{epsf}
%\epsfxsize=8.0cm
%\epsffile{burko/test.eps}
\centerline{\epsfysize 7.0cm
\epsfbox{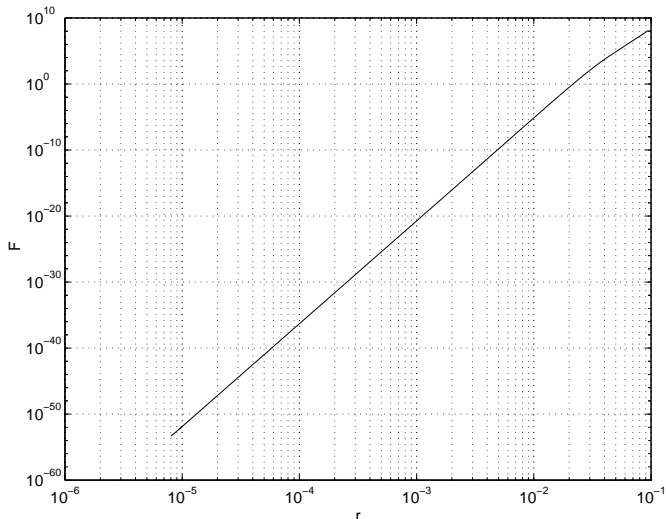}}
%\epsfbox{test.eps}}
\caption{Metric function $F=-2g_{uv}$ as a function of 
$r$ near the spacelike singularity along an outgoing null ray.}
\label{fig2}
\end{figure}

\begin{figure}
\input{epsf}
%\epsfxsize=8.0cm
%\epsffile{burko/sin2.eps}
\centerline{\epsfysize 8.0cm
\epsfbox{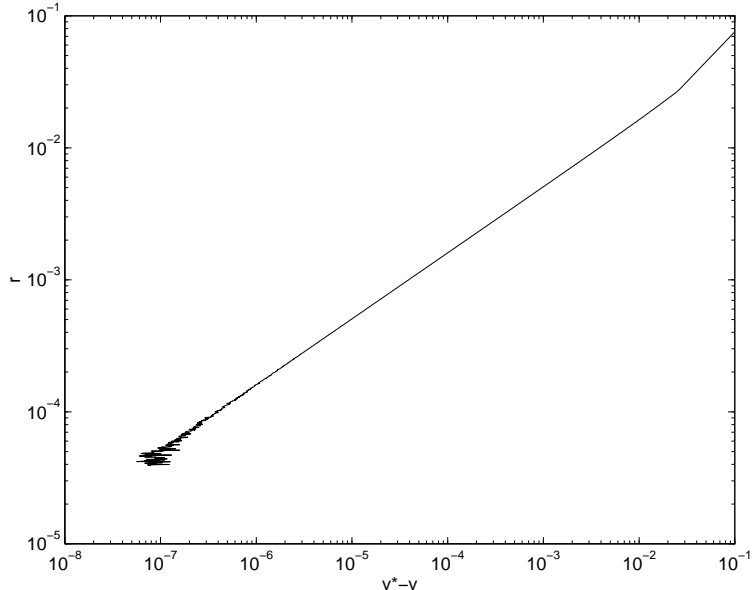}}
%\epsfbox{sin2.eps}}
\caption{Area coordinate $r$ as a function of $v_*-v$ near the 
spacelike singularity along an outgoing null ray. At small values of $r$ there is growing 
numerical noise. However, even 
before the noise becomes significant we have 
over two decades in $r$ and four decades in $v_*-v$ of asymptotic behavior.}
\label{fig3}
\end{figure}

This high amplitude scalar field increases the external mass of the black hole at late 
times by over $10\%$. We then probe the metric functions and the scalar 
field in the black hole's interior along outgoing (and ingoing) null rays, approaching 
the spacelike singularity, which is located at some finite value of the 
ingoing null coordinate $v_{*}(u_{p})$. Here, $u_{p}$ is the value of 
the outgoing null ray $u$ at which our outgoing null ray is located. 
Figure \ref{fig1} shows three such rays, denoted by 1,2, and 3. 
The metric we use for the numerics is 
\begin{equation}
\,ds^2=-F(u,v)\,du\,dv+r^2(u,v)\,d\Omega ^2.
\label{bur-met}
\end{equation}
The results we 
find numerically are as follow: First, the metric function 
$g_{uv}\equiv -F/2$  
vanishes at the singularity, and $F$ decays as a power-law 
with respect to the area coordinate $r$ as one gets closer to the 
singularity. In addition, we find that the area coordinate $r$ decays like 
a power of $v_{*}-v$ near the singularity. Numerically, we find in all 
our simulations this power to be 
very close to $\frac{1}{2}$, with a deviation 
of $1\%$. Therefore, we would like to show analytically, that in a 
generic solution indeed $r\propto (v_{*}-v)^{1/2}$ near the singularity. 
Our Numerical results are described in Figures \ref{fig2} and \ref{fig3}. These results are 
independent of the precise definition of the null coordinates we were 
using, as any regular gauge transformation which preserves the metric-form 
(\ref{bur-met}) introduces just a multiplicative factor which does 
not change the power-law indices. In addition,
we find similar results for all $u_{p}$'s. That is, the power with which 
$F$ decays to zero at the singularity depends on $u_{p}$, but 
only weakly so. The power $\frac{1}{2}$ we found for $r(v_{*}-v)$ is 
retained for all  $u_{p}$'s.

Therefore, we make the simplifying assumption 
of quasi homogeneity. Namely, we assume that only derivatives of the metric 
functions and of the scalar field normal to the singularity are 
non-vanishing, while derivatives tangent to the singularity are exactly 
zero on it. We thus write the line-element as 
\begin{equation}
\,ds^2=h(r)\,dt^2+f(r)\,dr^2+r^2\,d\Omega ^2,
\label{bur-met1}
\end{equation}
where $\,d\Omega ^2=\,d\theta^2+\sin^2\theta\,d\phi^2$ is the usual metric on the 
unit two-sphere. 
As the singularity is spacelike, and $t$ is the spacelike 
coordinate near the singularity and $r$ is timelike, we are going to 
neglect all derivatives with respect to $t$, and keep only derivatives 
with respect to $r$. With these assumptions the $t$--$t$, $r$--$r$ and 
$\theta$--$\theta$ components of the Einstein-Maxwell-scalar 
equations in spherical symmetry are, correspondingly,  
\begin{equation}
\frac{h}{r^2f^2}\left(f'r+f^2-f\right)=\frac{h}{f}\Phi '^2+h\frac{q^2}{r^4}
\label{bur-eq1}
\end{equation}
\begin{equation}
\frac{1}{hr^2}\left(h'r-hf+h\right)=\Phi '^2-f\frac{q^2}{r^4}
\label{bur-eq2}
\end{equation}
\begin{equation}
\frac{r}{4f^2h^2}\left(2h'fh-2f'h^2+2rh''hf-rh'^2f-rh'f'h\right)=
\frac{q^2}{r^2}-\frac{r^2}{f}\Phi'^2.
\label{bur-eq3}
\end{equation}
[The $\phi$--$\phi$ component of the field equations gives again 
Eq. (\ref{bur-eq3})]. 
Here, a prime denotes differentiation with respect to $r$. In addition 
we also have the Klein-Gordon equation for the scalar field 
$\Phi^{;\alpha}_{\;\;\; ;\alpha}=0$, whose first integral reads 
\begin{equation}
\Phi '(r)=\frac{d}{\sqrt{-g}}f(r)\;\sin\theta,
\label{bur-kg}
\end{equation}
where $d$ is an integration constant and $g$ is the metric determinant. (Note that 
because the metric determinant has a factor $\sin^2\theta$, $\Phi '(r)$ does not depend on $\theta$.)

We now use Eq. (\ref{bur-kg}) to eliminate the scalar field from Eqs. 
(\ref{bur-eq1})--(\ref{bur-eq3}), and find that 
\begin{equation}
f'r+f^2-f+\frac{f^2}{hr^2}\left(d^2-q^2h\right)=0
\label{bur-e1}
\end{equation}
\begin{equation}
h'r-hf+h+\frac{f}{r^2}\left(d^2+q^2h\right)=0
\label{bur-e2}
\end{equation}
\begin{equation}
h''-\frac{1}{2}\;\frac{h'^2}{h}+\frac{1}{r}\left(h'-\frac{f'}{f}h\right)
-\frac{1}{2}\;\frac{h'f'}{f}-2\frac{f}{r^4}\left(d^2+q^2h\right)=0.
\label{bur-e3}
\end{equation}
In what follows, we shall solve Eqs. (\ref{bur-e1}) and 
(\ref{bur-e2}), subject 
to our conclusions above from the numerical simulations. We then use 
Eq. (\ref{bur-e3}) as a consistency check for our solution. Namely, we 
seek a generic solution for which both metric functions $h,f$ vanish at 
the spacelike singularity. This is motivated by the vanishing of 
$g_{uv}$ in the double-null metric, from which the vanishing of both 
$h$ and $f$ follows, by merit of the relations  
$$g_{tt}\equiv h=2\frac{\,\partial u}{\,\partial t}
\frac{\,\partial v}{\,\partial t}g_{uv}=-2g_{uv}$$
$$g_{rr}\equiv f=g_{r_*r_*}\left(\frac{\,dr_*}{\,dr}\right)^2=
-g_{tt}\left(\frac{\,dr_*}{\,dr}\right)^2= 
2g_{uv}\left(\frac{\,dr_{*}}{\,dr}\right)^2.$$ Here, the 
`tortoise' coordinate $r_{*}$ is defined by $g_{r_{*}r_{*}}=-g_{tt}$, and 
the null coordinates are defined by $u=r_{*}-t$ and $v=r_{*}+t$.  

\section{Properties of power-law spacelike singularities}\label{sec3}
In this section, we shall study several properties of spacelike 
singularities, which are common to any spatially 
homogeneous metric of the form (\ref{bur-met1}) 
with $h(r),f(r)$ being powers of $r$. (Note, we still do not assume 
that $f(r),h(r)$ vanish at the singularity. We shall specialize to this 
case below.) First, we shall give a few examples of known 
spacelike singularities. Then, we shall study the general homogeneous 
power-law metric (\ref{bur-met1}). 

\subsection{Examples for spacelike singularities}
\subsubsection{The Schwarzschild singularity}
The Schwarzschild solution assumes near the singularity the form
\begin{equation}
\,ds^2\approx \left(\frac{r}{2M}\right)^{-1}\,dt^2-
\left(\frac{r}{2M}\right)\,dr^2+r^2\,d\Omega ^2,
\end{equation}
where $t$ is a spacelike coordinate tangent to the singularity and $r$ 
is timelike and normal to the singularity. This metric obviously 
satisfies the power-law form of (\ref{bur-met1}). 

The Kretschmann scalar along radial $t=\rm{const}$ 
curves is
\begin{equation}
R_{\alpha\beta\gamma\delta}R^{\alpha\beta\gamma\delta}=
\frac{64}{27}\,\frac{1}{(\tau_{*}-\tau)^4},
\label{bur-sch1}
\end{equation}
where $\tau$ is the proper time of the infalling object, 
and $\tau_{*}$ is the proper time at which the object arrives 
at the singularity at $r=0$. We now denote by $x^4$ the coordinate 
tanget to the world line of an infalling observer. Because the 
Schwarzschild spacetime is vacuous, the projection of the Ricci tensor 
on the world line is identically zero, i.e., 
$R_{(4)(4)}\equiv R_{\mu\nu}e^{\mu}_{(4)}e^{\nu}_{(4)}=0$, where 
$R_{(\alpha )(\beta )}$ is the $\alpha\beta$ tetrad component of the Ricci tensor. 
These results are independent of the mass $M$ 
of the Schwarzschild black hole. 

\subsubsection{The Friedmann-Robertson-Walker cosmology} 
The metric for the FRW cosmological model is given by 
\begin{equation}
\,ds^2=-dt^2+R^2(t)\left(\frac{1}{1-kr^2}\,dr^2+r^2\,d\Omega ^2\right).
\end{equation}
Let us consider here radiation-dominated perfect fluid with energy momentum 
tensor $T_{\mu\nu}=p\,g_{\mu\nu}+(p+\rho)u_{\mu}u_{\nu}$, $u_{\mu}$ 
being the four-velocity. We take $p=\rho /3$. 

Then, in a co-moving frame near the singularity 
\begin{equation}
R_{\alpha\beta\gamma\delta}R^{\alpha\beta\gamma\delta}=
\frac{3}{2}\,\frac{1}{(\tau_{*}-\tau)^4},
\end{equation}
and
\begin{equation} 
R_{(4)(4)}(\tau)=\frac{3}{4}\,\frac{1}{\tau ^2}.
\end{equation}
These results are independent of the density $\rho$ (or the pressure $p$), 
and are also independent of $k$, i.e., independent of the spatial 
curvature. 

\subsubsection{Kasner solution}
The Kasner solution with a spacelike parameter (similar results are 
obtained also for the Kasner solution with a timelike parameter) is given 
by the metric
\begin{equation}
\,ds^2=x^{2p_1}\,dt^2-\,dx^2-x^{2p_2}\,dy^2-x^{2p_3}\,dz^2,
\end{equation}
with $p_1+p_2+p_3=1=p_1^2+p_2^2+p_3^2$. (This solution is anisotropic and 
is not spherically symmetric like the other solutions we consider in 
this paper.) 

For the Kasner solution one finds 
\begin{equation}
R_{\alpha\beta\gamma\delta}R^{\alpha\beta\gamma\delta}=
16p_1^2(1-p_1)\,\frac{1}{(\tau_{*}-\tau)^4},
\end{equation}
and
\begin{equation} 
R_{(4)(4)}(\tau)=0.
\end{equation}
This last result is obvious from the fact that the Kasner solution 
represents a vacuum spacetime.

\subsection{General properties}
In all the above examples, the Kretschmann scalar diverges near the 
spacelike singularity like $1/(\tau_{*}-\tau)^4$, and the Ricci tensor, 
if spacetime is non-vacuum, diverges like $1/\tau ^2$. In this section 
we shall show that these results are general for any metric of the form 
(\ref{bur-met1}). 
Writing $h(r)=Br^m$ and $f(r)=-Ar^n$,  with $A,B$ being constants, 
a straightforward (but lengthy) calculation for the 
Kretschmann scalar yields
\begin{equation}
R_{\alpha\beta\gamma\delta}R^{\alpha\beta\gamma\delta}=
4\,\frac{m^4-4m^3-2m^3n+12m^2+4m^2n+m^2n^2+8n^2+16}
{(2+n)^4\,(\tau_{*}-\tau)^4}.
\label{bur-kre}
\end{equation}
In Schwarzschild, $m=-1$ and $n=1$. Substitution recovers the known result 
for Schwarzschild (\ref{bur-sch1}). This result (\ref{bur-kre}) is independent of $A,B$. 
This is the analogue of the independence of the Schwarzschild 
Kretschmann scalar of the mass $M$. One also finds that 
\begin{equation}
R_{(4)(4)}(\tau)=\frac{2m+mn+4n-m^2}{(n+2)^2}\,\frac{1}{\tau ^2}.
\label{bur-ric}
\end{equation}
This result is of great importance, because one can show that  
if $2m+mn+4n-m^2\ne 0$ than a 
sufficient condition for the singularity to be strong in the sense of 
Tipler is satisfied. 

A strong singularity in the sense of Tipler can be defined as follows 
\cite{bur-tip77}: 
Let us consider non-spacelike geodesic running into the singularity. 
Then, if the limit of the greatest lower bound 
of the volume element defined by the spatial 
metric induced by any three independent Jacobi fields vanishes as one 
approaches the singularity along every such geodesic, then the singularity 
is strong in the sense of Tipler. The physical content of this definition 
is that the volume of every extended physical object is compressed 
infinitely, such that every extended object will inevitably be crushed to zero volume. 
We note, that this is a `strong' definition for destructive singularities, 
as even weaker singularities would be strong enough to destroy any extended 
physical object. Tipler \cite{bur-tip77} gives an example for such a singularity, where 
the spatial metric induced by the Jacobi fields behaves near the 
singularity like ${\rm diag}(\tau ,\;\tau,\; \tau^{-2})$. The volume 
element defined by the spatial metric is unity, but any extended 
object would be infinitely stretched in one direction, and infinitely 
squeezed in two directions, in an infinite spaghettification. 

A theorem by Clarke and Kr\'{o}lak \cite{bur-cla85} 
relates $R_{(4)(4)}$ to the strength of the singularity: A 
sufficient condition for a singularity to be strong in the Tipler 
sense is that the integral 
$\int_0^{\tau_{*}}\,d\tau ' \int_0^{\tau '}\,d\tau ''R_{(4)(4)}(\tau '')$ 
diverges at $\tau_{*}$. For example, the FRW cosmological model has 
a Tipler strong singularity, as for this case this integral diverges 
logarithmically. Note that this is a sufficient condition, but not a 
necessary condition: the Schwarzschild and Kasner singularities are 
Tipler strong, but the integral vanishes identically by merit of their 
vacuous spacetimes. The important conclusion to be made here is as follows: 
For any non-vacuum 
homogeneous spacelike singularity near which the metric has the 
power-law form (\ref{bur-met1}) the singularity is strong in the Tipler 
sense by virtue of the theorem by Clarke and Kr\'{o}lak and by Eq. (\ref{bur-ric}). 

\section{The uncharged case}\label{sec4}
In this section we study the spacelike singularity in the presence of a 
self-gravitating scalar field with spherical symmetry and with vanishing electric 
charge $(q=0)$. This is the non-linear 
generalization in the homogeneous case of the linear analysis of 
Doroshkevich and Novikov \cite{bur-dor78}. In the next section we shall 
also study the charged case. We assume a series expansion for the metric functions 
of the general form 
$f^{(n)}(r)=\sum_{i=1}^{n}f_i(r)$ and $h^{(n)}(r)=\sum_{i=1}^{n}h_i(r)$, where 
at $r=0$ $f^{(n)}=h^{(n)}=0$. Here, $n$ denotes the expansion order of the 
series, which, as will be shown below, are assumed to have a finite radius 
of absolute convergence. That is, $f^{(n=\infty)}\equiv f$, $h^{(n=\infty)}\equiv h$ 
for some finite interval $0\le r <r_0$. 
We next assume a power-law series behavior of the metric functions. 
That is, we assume that $h_i=h_{i0}\;r^{n_i}$ and $f_i=f_{i0}\;r^{m_i}$.  
Note that for the time being we do not restrict the parameter 
range of $n_i ,\;m_i$. 

\subsection{The leading order approximation}
 
To the leading order all nonlinear terms in the rhs of Eqs.  (\ref{bur-e1}) and 
(\ref{bur-e2}) are negligible, and therefore Eqs. (\ref{bur-e1}) and (\ref{bur-e2}) assume the form 
\begin{equation}
f_1'r-f_1+\frac{d^2f_1^2}{h_1r^2}=0
\label{bur-o11}
\end{equation}
\begin{equation}
h_1'r+h_1+\frac{d^2f_1}{r^2}=0
\label{bur-o12}
\end{equation}
Substitution of the Ansatz $h_1=B\;r^{\beta}$, $f_1=-A\;r^{\alpha}$, with $A,B>0$ 
(as we are looking for a spacelike singularity) 
into these equations yields $\alpha =\beta +2$ and 
$A=(\beta +1)B/d^2$. From the positivity of $A$ and $B$ it then follows that 
$\beta >-1$. In Schwarzschild, $\beta =-1$, and the exponents are uniquely 
determined. Here, however, the scalar field endows the field equations with a 
freedom in the value of $\beta$. For small amplitudes of the scalar field on the 
initial hypersurface we would expect the deviation of $\beta$ from $-1$ to be 
small, such that $\beta$ would still be negative. However, 
there would also be situations with positive values for 
$\beta$, and even $\beta=0$ as a special case. 
We thus find that 
\begin{equation}
h^{(1)}=B\;r^{\beta}
\label{bur-o13}
\end{equation}
\begin{equation}
f^{(1)}=-(\beta +1)\;\frac{B}{d^2}\;r^{\beta +2}.
\label{bur-o14}
\end{equation}
Recall that $d$ is a constant of integration for the Klein-Gordon equation (\ref{bur-kg}). In 
Schwarzschild $d=0$, and therefore one needs to take 
$$\lim_{\begin{array}{c}d\rightarrow 0\\ \beta\rightarrow -1\end{array}}
\frac{\beta +1}{d^2}=\frac{1}{4M^2}\;\;\;\;\;\;\;\;\;\;\;\;\;
\lim_{\begin{array}{c}d\rightarrow 0\\ \beta\rightarrow -1\end{array}}B=2M,$$
where $M$ is the mass of the black hole, in order to recover  
in the limit of vanishing scalar field  
the Schwarzschild solution.  
An important conclusion to be drawn from this solution is that 
$r\propto (v_{*}-v)^{1/2}$. This can be shown as follows: 
For a general homogeneous metric (\ref{bur-met1}) 
we define a `tortoise' coordinate $r_{*}$ by $g_{r_*r_*}=-g_{tt}$. The metric then takes the form 
\begin{equation}
\,ds^2=g_{r_*r_*}\left(\,dr_*^2-\,dt^2\right)+r^2\,d\Omega ^2,
\end{equation}
where $g_{r_*r_*}=-g_{tt}=g_{rr}\;\left(\,dr/\,dr_*\right)^2$. 
\newline
Consequently, 
$\,dr/\,dr_*=-\sqrt{-g_{tt}/g_{rr}}=-(1/r)\sqrt{d^2/(\beta +1)}$. Integration yields 
$r_*=-\frac{1}{2}\sqrt{(\beta +1)/d^2}\;r^2+{\rm const}$. Defining now future-directed null coordinates by 
$t=\frac{1}{2}(v-u)$ and $r_*=\frac{1}{2}(v+u)$, we find 
$v=2r_*+{\rm const}$, or 
\begin{equation}
r=\left(\frac{d^2}{\beta +1}\right)^{\frac{1}{4}}\;\left(v_*-v\right)^{1/2}.
\label{bur-rsqrtv}
\end{equation}
The power $\frac{1}{2}$ we find is independent of the value of $\beta$, 
and is a direct consequence of the relation $\alpha =\beta +2$. We note that for 
Schwarzschild $\alpha=1$ and $\beta=-1$, which clearly satisfies this relation. 

We note, that the square-root relation of Eq. (\ref{bur-rsqrtv}) can be deduced directly 
from the full partial differential equations. From our numerical results $F(u,v)$ vanishes 
faster than $r(u,v)$ as functions of $v-v_{*}$ near the singularity. 
Consequently, from Eq. (6) of Ref. \cite{bur-bur97} it follows that near the singularity 
$(r^2)_{,uv}\approx 0$. This equation can be readily integrated to 
$r^{2}(u,v)=x_{1}(u)+x_{2}(v)$. As at the singularity $r$ vanishes, one finds that 
along $u_p={\rm const}$ a series expansion yields 
$$\left.r^2(v)\right|_{u_p}=(v-v_{*})\;\left.\frac{\,d(r^2)}{\,dv}\right|_{v_{*}}+O[(v-v_{*})^2],$$ 
and consequently $r\propto (v-v_{*})^{1/2}$.

We next show that our solution is generic. 
In our solution there are three arbitrary parameters, 
namely $\beta$, $d^2$, and $B$. 
Without a scalar field the solution is Schwarzschild, 
where there is just one arbitrary parameter, by virtue of Birkhoff's theorem. 
Therefore, with the scalar field, we would expect one additional arbitrary parameter, i.e., 
two arbitrary parameters. 
Apparently, we have here three parameters. 
However, the arbitrariness in the fixing of $B$ is just a trivial 
gauge mode, related to the possibility to make an arbitrary 
transformation $t\to t'=f(t)$ ({\it cf}. Ref. \cite{bur-lan75} for the 
analogue in Schwarzschild). 
Therefore, with a scalar field there are only two non-trivial 
arbitrary parameters, as should be expected. Therefore, our solution has the right 
number of arbitrary parameters, and in this sense is generic. Now, for a 
system of nonlinear equations the notion of a general solution is not 
unambiguous, and there may be other solutions of non-zero measure in 
solutions space, but our solution above also has a non-zero measure, and 
is therefore generic.

We can now substitute our solution in the Klein-Gordon 
equation (\ref{bur-kg}), and obtain for the scalar field to the leading order in $r$ 
\begin{equation}
\Phi ^{(1)}=\sqrt{\beta +1}\;\ln r
\label{bur-so11}
\end{equation}
which diverges logarithmically like the linear 
scalar field studied by Doroshkevich and Novikov \cite{bur-dor78}. 
We note that the amplitude of the scalar field depends on $\beta$, which is 
determined by the nonlinear equations, whereas in the linear analysis 
the amplitude in arbitrary. Therefore, even the leading order in 
the expansion for the scalar field is determined by nonlinear effects. 

We now assess the error involved with the consideration of the leading order 
only of the series expansion. For this aim, we consider Eqs. 
(\ref{bur-e1})--(\ref{bur-e3}). Substituting our leading order expression, we find the 
expressions for the deviation from zero of the rhs of each equation, which we shall refer 
to below as the error associated with the truncated solution at a certain order.  
We find that Eq. (\ref{bur-e3}) is satisfied exactly. To the leading order in $r^{\beta}$ 
the error in Eq. (\ref{bur-e1}) is $(B^2/d^4)\;(\beta +1)^2\;r^{2\beta +4}$, and 
the error in Eq. (\ref{bur-e2}) is $(B^2/d^2)\;(\beta +1)\;r^{2\beta +2}$. 

\subsection{The second order approximation}
We shall now find the second order in the series expansions for the metric 
functions and for the scalar field.  To the second order the field equations 
Eq. (\ref{bur-e1}) and Eq. (\ref{bur-e2}) reduce to
\begin{equation}
f'_2r-\left(1-2\frac{d^2f_1}{h_1r^2}\right)\;f_2+\left(1-\frac{d^2h_2}
{r^2h_1^2}\right)\;f_1^2=0
\label{bur-o21}
\end{equation}
\begin{equation}
h'_2r+h_1f_1+h_2+\frac{d^2f_2}{r^2}=0,
\label{bur-o22}
\end{equation}
where $h_1$ and $f_1$ are already known from Eqs. (\ref{bur-o13}) and 
(\ref{bur-o14}). We again assume a power-law 
behavior for $f_2$ and $h_2$, and obtain
\begin{equation}
f_2=-\frac{(\beta +1)^2(3\beta +4)}{(\beta +2)^2}\;\frac{B^2}{d^4}\;r^{2\beta +4}
\label{bur-o23}
\end{equation}
\begin{equation}
h_2=\frac{\beta (\beta +1)}{(\beta +2)^2}\;\frac{B^2}{d^2}\;r^{2\beta +2}.
\label{bur-o24}
\end{equation}
We thus find that the second-order expansion for $f$ and $h$ is
\begin{equation}
f^{(2)}=-(\beta +1)\frac{B}{d^2}\;r^{\beta +2}
-\frac{(\beta +1)^2(3\beta +4)}{(\beta +2)^2}\;\frac{B^2}{d^4}\;r^{2\beta +4}
\end{equation}
\begin{equation}
h^{(2)}=B\;r^{\beta}+\frac{\beta (\beta +1)}{(\beta +2)^2}\;\frac{B^2}{d^2}\;r^{2\beta +2}.
\end{equation}

We now evaluate the errors, to leading order in $r^{\beta}$, 
in the field equations (\ref{bur-e1})--(\ref{bur-e3}) 
for the second-order expansion. For Eq. (\ref{bur-e1}) the error is  
$2(B^3/d^6)\;[(\beta +1)^3(5\beta +6)/(\beta +2)^2]\;r^{3\beta +6}$, 
the error in Eq. (\ref{bur-e2}) is $4(B^3/d^4)\;[(\beta +1)^3/(\beta +2)^2]
\;r^{3\beta +4}$, and the error in Eq. (\ref{bur-e3}) is
$[(\beta +1)(\beta^2+8\beta+8)/(\beta+2)^2]/d^4\;r^{3\beta+2}$.

The second order expansion we find from the Klein-Gordon equation (\ref{bur-kg}) 
for the scalar field is
\begin{equation}
\Phi ^{(2)}=\sqrt{\beta +1}\;\ln r+
\frac{\sqrt{\beta +1}\;(5\beta^2+12\beta+6)}{(\beta+2)^3}\;\frac{B}{d^2}\;r^{\beta+2}.
\end{equation}
Namely, under nonlinear effects the powers of the second (and higher) 
order term is different from the power found in the linear analysis \cite{bur-dor78}. 

\subsection{General expression for the series expansion}
We now turn to the general form of the series expansion. Based on the above 
expressions, we seek series expansions of the forms
\begin{equation}
f=\sum_{n=1}^{\infty}a_n\;r^{(\beta+2)n}\;\;\;\;\;\;\;\;\;\;\;
a_n=a_n(d^2\;,\beta\;,B)
\label{bur-se1}
\end{equation}
\begin{equation}
h=\sum_{n=1}^{\infty}b_n\;r^{(\beta+2)n-2}\;\;\;\;\;\;\;\;\;
b_n=b_n(d^2\;,\beta\;,B).
\label{bur-se2}
\end{equation}
Equations (\ref{bur-se1}) and (\ref{bur-se2}) will solve the field equations 
(\ref{bur-e1})--(\ref{bur-e3}) if the following two conditions are satisfied: 
First, the series converge absolutely [this is to ensure that when Eqs. 
(\ref{bur-se1}) and (\ref{bur-se2}) are substituted in Eqs. 
(\ref{bur-e1})--(\ref{bur-e3}) the series multiplication theorem will be applicable], 
and second, that the expansion coefficients $a_n$ and $b_n$ can be found 
uniquely for any $n$. If these two conditions are satisfied, than Eqs. 
(\ref{bur-se1}) and (\ref{bur-se2}) represent a generic solution of the field equations 
(as they contain the right number of arbitrary parameters). The first condition is hard to 
prove, without knowledge of the values of the arbitrary functions $d^2$, 
$\beta$ and $B$. We thus assume that the first 
condition is satisfied for small enough values of $r$. 
We next prove the satisfaction of the second condition. 
For $n\ge 2$, Eqs. (\ref{bur-se1}) and (\ref{bur-se2}) yield the following 
algebraic equations for the expansion coefficients $a_n$ and $b_n$:
\begin{equation}
d^2\;a_n+\left[(\beta+2)(n+1)-1\right]\;b_n=\sum_{k=0}^{n-1}a_{n-k-1}b_k
\end{equation}
\begin{equation}
\left\{\left[n(\beta+2)+\beta+1\right]\;b_1+2d^2\;a_1\right\}\;a_n+(\beta+1)a_1\;b_n=c,
\end{equation}
where 
\begin{eqnarray*}
c=&-&\sum_{k=1}^{n-1}\left\{\left[(n-k)(\beta+2)
\;b_k+d^2\;a_k\right]\;a_{n-k}+b_k\sum_{l=0}^{n-k-1}a_{l}a_{n-k-l-1}\right\}\\
&-&b_1\sum_{l=0}^{n-1}a_{l}a_{n-l-1}.
\end{eqnarray*}
These equation will have a unique solution if the determinant of the 
homogeneous part does not vanish for all $n$. A straightforward 
calculation yields for this determinant $\Delta_n=n^2B(\beta+2)^2\ne 0$ 
(recall that $\beta>-1$), which proves that the second condition is satisfied. 

The general series expansion for the scalar field is 
\begin{equation}
\Phi =\sqrt{\beta+1}\;\left(\ln r+\sum_{n=2}^{\infty}c_n\;r^{(\beta+2)(n-1)}\right),
\end{equation}
where $c_n$ are the expansion coefficient which can be found uniquely for each order 
$n$ from the Klein-Gordon equation (\ref{bur-kg}).
We note that in Schwarzschild $\beta=-1$, and indeed for this choice of $\beta$ the scalar field vanishes identically.

\section{The charged case}\label{sec5}
In this section we shall assume that $q\ne 0$, and show that there exists a 
generic solution for a spacelike singularity, of similar properties 
with the singularity one finds numerically \cite{bur-bra95,bur-bur97a}. We shall show that this singularity has 
many similarities 
to the singularity in the uncharged case we studied in the previous section, but also some important differences. 
We again make the Ansatz that the 
metric functions are functions of $r$ only, and that $r$ is a timelike coordinate.  
We thus neglect all derivatives with respect to spacelike coordinates.  
We first consider the leading order expression, 
then the second order correction, and finally the general expression for the series expansion.

\subsection{The leading order approximation}
The source term for the Einstein-Maxwell-Klein-Gordon equations Eqs. 
(\ref{bur-eq1})--(\ref{bur-eq3}) 
contains two contributions: the contribution of the electric field and the contribution of 
the scalar field. In order to find the leading order of the series expansion for the 
metric functions, one can consider the following three possibilities for the relative 
contributions of the two sources near the spacelike singularity: 
\newline
(a) The contribution of the scalar field near $r=0$ is dominant, and the electric field's contribution is negligible. 
\newline
(b) The contribution of the electric field near $r=0$ is dominant, and the scalar field's contribution is negligible.
\newline
(c) The contributions of the scalar field and of the electric field near $r=0$ are comparable.
\newline
It turns out that there is no consistent solution of the field 
equations with possibility (c). Let us now consider possibility (b): As near $r=0$ the scalar field's 
contribution to the field equations is negligible compared with the electric field's 
contribution, the leading order expression is the same as the leading order 
expression without a scalar field at all. However, from the generalized Birkhoff 
theorem, the solution then is nothing but the Reissner-Nordstr\"{o}m solution, written as a series 
expansion near $r=0$. But the 
$r=0$ singularity in Reissner-Nordstr\"{o}m is timelike rather than 
spacelike, which is the type of singularity we are interested at here. 
Therefore, we do not expect possibility (b) to realize in our case. We note in passing 
that this case might be of some relevance for consideration of the self-gravitating 
scalar field solution for an hypothetical 
extension of the spacetime manifold {\it beyond} the weakly singular Cauchy horizon.

We study, then, possibility (a). Namely, we assume that near the singularity
the contribution of the electric field to the energy-momentum tensor is
negligible compared with the contribution of the scalar field. With this
assumption, the equations which govern the first order expressions for the
metric functions and for the scalar field are the same as their uncharged
counterparts Eq. (\ref{bur-o11}) and Eq. (\ref{bur-o12}). Consequently, 
to the leading order, the spacelike singularities
with an electric field or without an electric field have the same functional
form [{\it cf.} Eqs. (\ref{bur-o13}) and (\ref{bur-o14})]:
\begin{equation}
h^{(1)}=B\;r^{\beta}
\label{bur-o15}
\end{equation}
\begin{equation}
f^{(1)}=-(\beta +1)\,\frac{B}{d^2}\;r^{\beta +2}.
\label{bur-o16}
\end{equation}
There is, however, a difference in the range of the parameters between
the two cases. Whereas in the uncharged case $\beta$ was free to assume
negative, positive or zero values (subject to the restriction
$\beta >-1$), we here find that in order to have a solution consistent
with our conclusions based on numerical simulations $\beta$ needs to be
positive. Without a scalar field, the solution is Reissner-Nordstr\"{o}m,
where the singularity is timelike, and $\beta =-2$. Because there is no
scalar field, the electric contribution to the energy-momentum tensor cannot
be neglected, and the first order equations are dominated by the electric
contribution. Therefore, there is no smooth analytic limit for vanishing
scalar field, and there is a discontinuity in the values that $\beta$
can take. This can be understood in terms of the causal structure of the
singularity, which with a scalar field is spacelike and without a scalar
field is timelike. With Eqs. (\ref{bur-o15}) and (\ref{bur-o16}), 
the Klein-Gordon equation (\ref{bur-kg}) yields for the scalar field again 
\begin{equation}
\Phi ^{(1)}=\sqrt{\beta +1}\;\ln r
\label{bur-so12}
\end{equation}
which is the same as $\Phi^{(1)}$ for the uncharged case [Eq. (\ref{bur-so11})]. Note 
that there is no value of $\beta$ which nullifies the scalar field in this case, 
as with vanishing scalar field there is no spacelike singularity like the singularity we are studying. 

The estimate for errors associated with the leading order terms in the charged case are, 
to the leading order in $r^{\beta}$: For Eq. (\ref{bur-e1}) the error is 
$-(B^2/d^4)\;(\beta +1)^2\;q^2\;r^{2\beta +2}+(B^2/d^4)\;(\beta +1)^2 
r^{2\beta +4}$, for Eq. (\ref{bur-e2}) the error is 
$-(B^2/d^2)\;(\beta +1)\;q^2\;r^{2\beta}+(B^2/d^2)\;(\beta +1)\;r^{2\beta+2}$, 
and for Eq. (\ref{bur-e3}) the error is 
$2(B^2/d^2)\;(\beta+1)\;q^2\;r^{2\beta -2}$. 
Note the differences between these error estimates and their counterparts in the uncharged case. 
For each equation, the error estimate here has two terms, one of them does not depend on the charge, and the 
other does depend on it. The term which does not depend on the charge is the same 
as the error term in the uncharged case. However, 
the charge-dependent term has a smaller power index, 
and is therefore dominant. [In the error for Eq. (\ref{bur-e3}) the term 
which does not depend on the charge vanishes.] The difference in the 
power indices between the charge-dependent and the charge-independent 
terms is $2$. We shall show below that this is an indication for a more complicated series form for the metric 
functions than the form we encountered in the uncharged case. 

\subsection{The second order approximation}
The electric field does not contribute then to the leading 
order terms in the series expansion for the metric 
functions or the scalar field. However, we find that it does 
contribute to the second order terms. The equations which govern the second order terms are:
\begin{equation}
f_2'r-\left(1-2\frac{d^2f_1}{h_1r^2}\right)\;f_2-\left(\frac{q^2}{r^2}+
\frac{d^2h_2}{r^2h_1^2}\right)\;f_1^2=0
\end{equation}
\begin{equation}
h_2'r+h_2+\frac{d^2f_2}{r^2}+q^2\;\frac{h_1f_1}{r^2}=0,
\end{equation}
where again $f_1$ and $f_2$ are known from the first order terms (\ref{bur-o15}) and 
(\ref{bur-o16}), and where we again assumed a power-law behavior. 
These equations are different from their uncharged counterparts [Eqs. (\ref{bur-o21}) and (\ref{bur-o22})] 
not only in the appearance of charge-dependent terms, 
but also in the absence of terms which appeared in the uncharged case. 
Substitution of the power-law behavior Ansatz yields
\begin{equation}
f_2=\frac{(\beta+1)^2(3\beta+2)}{\beta ^2}\;\frac{B^2}{d^4}\;q^2\;r^{2\beta +2}
\end{equation}
\begin{equation}
h_2=-\frac{(\beta+1)(\beta+2)}{\beta ^2}\;\frac{B^2}{d^2}\;q^2\;r^{2\beta},
\end{equation}
such that
\begin{equation}
f^{(2)}=-(\beta +1)\,\frac{B}{d^2}\;r^{\beta +2}+
\frac{(\beta+1)^2(3\beta+2)}{\beta ^2}\;\frac{B^2}{d^4}\;q^2\;r^{2\beta +2}
\end{equation}
\begin{equation}
h^{(2)}=B\;r^{\beta}
-\frac{(\beta+1)(\beta+2)}{\beta ^2}\;\frac{B^2}{d^2}\;q^2\;r^{2\beta}
\end{equation}
and from the Klein-Gordon equation (\ref{bur-kg})
\begin{equation}
\Phi^{(2)}=\sqrt{\beta+1}\;\ln r-\frac{\sqrt{\beta+1}\;(\beta+1)(3\beta+2)}
{2\beta^3}\;\frac{B}{d^2}\;q^2\;r^{\beta}.
\end{equation}

The estimate for the leading terms in the errors associated with the 
second order terms are: 
For Eq. (\ref{bur-e1}) the error is 
$(\beta +1)^2\;(B^2/d^4)\;r^{2\beta+4}+2[(\beta+1)^3(5\beta+4)/\beta^2]\;
(B^3/d^6)\;q^2\;r^{3\beta+2}+O(r^{3\beta+4})$, 
for Eq. (\ref{bur-e2}) the error is 
$(\beta+1)\;(B^2/d^2)\;r^{2\beta+2}+4[(\beta+1^3)/\beta^2]\;(B^3/d^4)\;q^4\;r^{3\beta}+
O(r^{3\beta+4})$, 
and for Eq. (\ref{bur-e3}) the error is
$[(\beta+1)^2(\beta^2-6\beta-8)/\beta^2]\;(B^3/d^4)\;q^4\;r^{3m-2}
+O(r^{4\beta-2})$. 

Comparing these
expressions for the errors associated with the second order expansions and the errors associated
with the first-order expansions reveals that the effect of the second-order terms was to provide
terms which balance the leading-order errors for the first-order expansions, and leave just the subsequent
terms of the first-order expansions as the leading terms for the errors of the second-order expansions. 
Namely, the contribution to the solution of each order balances the leading-order terms in the errors 
associated with the lower-order solution, and thus reduces the errors.

\subsection{General expression for the series expansion}
One would be tempted to consider a series form similar to the series 
form we have for the uncharged case (\ref{bur-se1}) and (\ref{bur-se2}). However, when one considers the third order 
terms for the metric functions $f_3$ and $h_3$ in the charged case 
one finds that this is impossible, unless the exact value of 
$\beta$ is known. A close inspection of our previous results 
and of the third order approximation shows 
that the increments in the power indices in the charged case 
are not equal. Namely, the increment changes from $\beta$ to 
$\beta+2$. Therefore, it is more natural to seek a 
general solution in terms of a double series expansion than a single series. Therefore, the solution will be of the form
\begin{equation}
f=\sum_{m=1}^{\infty}\sum_{n=1}^{\infty}a_{mn}\;r^{\beta(m-1)+(\beta+2)n}\;\;\;\;\;\;\;\;\;\;a_{mn}=a_{mn}(d^2,\;q^2,\;B,\;\beta)
\label{bur-se3}
\end{equation}
\begin{equation}
h=\sum_{m=1}^{\infty}\sum_{n=1}^{\infty}b_{mn}\;r^{\beta  m+(\beta+2)(n-1)}\;\;\;\;\;\;\;\;\;\;b_{mn}=b_{mn}(d^2,\;q^2,\;B,\;\beta).
\label{bur-se4}
\end{equation}
Again, in order that Eqs. (\ref{bur-se3}) and (\ref{bur-se4}) 
would be a solution of the field equations (\ref{bur-e1})--(\ref{bur-e3}), the series 
should converge absolutely for some finite convergence interval, and the series expansion coefficients 
$a_{mn}$ and $b_{mn}$ should be found uniquely for each $m,n$. In fact, the 
terms we previously found explicitly are nothing but the $a_{11}$, $a_{21}$ and 
$b_{11}$, $b_{21}$ terms in this double series expansion. It is still an open question 
whether the satisfaction of these two conditions can be proved 
rigorously. A similar double series expansion is also obtained for 
the scalar field, with the leading term diverging logarithmically. 

\section{Singularity strength}\label{sec6}
From the expression for the metric functions we found 
for the charged case we can calculate the leading term for the Kretschmann scalar and for $R_{(4)(4)}$ in terms 
of the proper time of an observer who follows a $t={\rm const}$ radial trajectory. 
We find that 
\begin{equation}
R_{\alpha\beta\gamma\delta}R^{\alpha\beta\gamma\delta}=
64\;\frac{2\beta (\beta+1)+3}{(\beta+4)^4}\;\frac{1}{(\tau_*-\tau)^4}
\end{equation}
\begin{equation}
R_{(4)(4)}(\tau )=8\;\frac{\beta+1}{(\beta+4)^2}\;\frac{1}{\tau ^2}.
\label{bur-ricci}
\end{equation}
From Eq. (\ref{bur-ricci}) and from the theorem by Clarke and Kr\'{o}lak 
\cite{bur-cla85} it then follows that the spacelike singularity 
in spherical charged black holes with a self-gravitating scalar field is strong in the Tipler sense. Recall that 
the singularity we found is a generic one, in the sense that it 
relies on the correct number of arbitrary parameters. 
One indeed finds that these are the expressions which one obtains 
from the general expressions for power-law metrics (\ref{bur-kre}) and (\ref{bur-ric}) for this 
solution with $m=\beta$ and $n=\beta+2$. In addition, as 
in the charged case $\beta >0$, it turns out that there is no special 
case which nullifies either of the curvature invariant 
$R_{\alpha\beta\gamma\delta}R^{\alpha\beta\gamma\delta}$ or the second integral over $R_{(4)(4)}(\tau )$. Namely, for any choice 
of $\beta$ the spacelike singularity is strong in the Tipler sense. 
(Note that because of the inevitable presence of the scalar field for the 
singularity to be spacelike with electric charge, there is no vacuum solution, and therefore the Ricci tensor does not vanish.)

\section{Concluding remarks}\label{sec7}
We found a generic solution for a spacelike singularity for the spherically symmetric Einstein-Maxwell-scalar field 
equations, which was previously found numerically in scalar-field collapse simulations.  The generic singularity 
we found has the same properties as the singularity which arises in the numerical simulations. 

The singularity we found was obtained under the assumption of quasi-homogeneity, namely, that the metric 
functions do not depend on derivatives with respect to the spatial coordinates 
(in our case, because of the spherical symmetry, this means that there 
is no dependence on $t$). Namely, we sought a velocity-dominated singularity. 
The next obvious step in the analysis of the singularity would be to relax the homogeneity assumption. 
This could perhaps be done by allowing the arbitrary parameters in the solution 
to be functions of $t$. Namely, all we need to do is to make the transformation 
$\beta\to\beta (t)$, $B\to B(t)$, and $d^2\to d^2(t)$ for the leading order terms of the solution. The 
higher-order terms will have to be corrected for the terms with dependence on derivatives with respect 
to $t$. (Because of the spherical symmetry there would still not be derivatives 
with respect to the angular coordinates $\theta$ and $\phi$.)  A preliminary check shows that these leading terms indeed satisfy the 
full spherically-symmetric Einstein-Maxwell-scalar field equations, 
with error smaller than the leading terms. However, it is still needed to perform a fuller analysis of the higher-order contributions. 
Then, the solution would be inhomogeneous, and depend on three 
arbitrary functions, and in this sense would be a generic solution. 
The homogeneous solution is then nothing but the pointwise behavior of the inhomogeneous solution. However, it is still to be 
shown that this inhomogeneous solution really solves the inhomogeneous Einstein-Maxwell-scalar field equations. 

Special attention is required in order to analyze to behavior of the fields near the spacetime event corresponding 
to the change of the causal structure of the singularity, namely to the event where the null singularity becomes 
spacelike. There are two possibilities to approach this spacetime event numerically, using a double-null code 
such as ours. First, one could 
approach this event along an outgoing null rays. Then one would need to reduce the value of the outgoing null coordinate 
$u$, and compare the 
values of the fields approaching the singularity along various values of $u$.
Second, one could also study the singularity along ingoing null rays. Late-time ingoing null rays actually probe 
the null mass-inflation singularity. However, one can also focus attention to the deep portions of these rays, and thus 
study the nearly-complete focusing domain, where the value of $r$ is very small.  
We find numerically that the area coordinate $r$ continues to decrease 
monotonically even at the nearly-complete focusing domain. 
This second approach would be more natural with a sleight variation of the numerical code: 
The integration is normally carried out in our code along ingoing rays, namely, we intergate from the 
initial data on an ingoing ray and the data on the 
first grid-point on the next ingoing ray, the fields throughout this second ray, and so on 
(for more details see Ref. \cite{bur-bur97}). 
However, for this second approach to the 
special event where the causal structure of the singularity changes, one could benefit from changing the 
direction of integration, and 
integrate along {\em outgoing} null rays instead of ingoing rays. Then, for each value of advanced time $v$ for 
a certain ingoing null ray 
one would eventually approach the spacelike singularity for large $u$, and the later $v$, 
the closer one would be the the spacetime event under consideration.

We again stress that it is still an open question whether in a more realistic model our results would be preserved. 
Namely, a scalar field was introduced as a toy model for a physical field because it 
has a radiative mode in spherical symmetry. However, a scalar field is not a 
realistic physical field, and it is possible that more realistic fields will 
not create a spacelike singularity. It has been shown before that scalar 
fields create unique phenomena (see, e.g., Ref. \cite{bur-bel73}), and therefore 
it is not impossible that the generic spacelike singularity we found is just an 
artifact of an over-simplified toy model. In addition, a realistic black hole will not be 
spherically symmetric, and whether a spacelike singularity would be created inside 
spinning black holes is a question still awaiting investigations.

\section*{Acknowledgements}

I thank Amos Ori for many stimulating discussions and Alexei Starobinsky for useful comments.

\end{document}